\documentclass[rnote]{aa}

\usepackage{graphicx}
\usepackage{txfonts}

\begin{document}
\title{The Star Pile in Abell 545\thanks{Based on observations obtained at the Gemini Observatory, which is 
operated by the Association of Universities for Research in Astronomy, 
Inc., under a cooperative agreement with the NSF on behalf of the Gemini 
partnership: the National Science Foundation (United States), the 
Science and Technology Facilities Council (United Kingdom), the National 
Research Council (Canada), CONICYT (Chile), the Australian Research 
Council (Australia), CNPq (Brazil) and CONICET (Argentina). Also based on
observations taken at the European Southern Observatory, Cerro Paranal, Chile;
ESO program 70.B-440(A).}}

\author{R. Salinas
           \inst{1}
           \and
        T. Richtler
           \inst{1}
           \and
        A. J. Romanowsky
         \inst{1,2}
         \and
         M. J. West
         \inst{3,4}
         \and
         Y. Schuberth
         \inst{5,1}     
}

\offprints{R. Salinas.}

\institute{Departamento de F\'{\i}sica, Casilla 160-C, Universidad de Concepci\'on, Concepci\'on, Chile\\
                \email{rsalinas@astro-udec.cl, tom@mobydick.cfm.udec.cl}
        \and
        UCO/Lick Observatory, University of California, Santa Cruz, CA 95064,
USA\\
\email{romanow@ucolick.org}
 \and
Gemini Observatory, Casilla 603, La Serena, Chile
\and
European Southern Observatory, Alonso de C\'ordova 3107, Vitacura, Casilla 19001, Santiago 19, Chile \\
\email{mwest@eso.org}
\and   
      Argelander Institut f\"ur Astronomie, Auf dem H\"ugel 71, 53121 Bonn, Germany\\
            \email{yschuber@astro.uni-bonn.de}
}
\date{}

\abstract{Struble (1988) found what appeared to be a cD halo without cD
  galaxy in the center of the galaxy cluster Abell 545. This remarkable
  feature has been passed almost unnoticed for nearly twenty years.}
        {Our goal is to review Struble's claim by providing a first
          (preliminary) photometric and spectroscopic analysis of this ``star pile''.}
{Based on archival VLT-images and long-slit spectra obtained with Gemini-GMOS, we describe the photometric structure and measure the redshift of the star pile and of the central galaxy.}{The star pile is indeed associated with Abell 545. Its
velocity is higher by about 1300 km/s than that of the central object. The spectra indicate an old, presumably metal-rich
population. Its brightness profile is much shallower than that of typical cD-galaxies.}{The formation history and the
dynamical status of the star pile remain elusive, until high S/N spectra and a dynamical analysis of the galaxy cluster
itself become available. We suggest that the star pile might provide an interesting
test of the Cold Dark Matter paradigm.}

\keywords{Galaxies: clusters: individual: A545 - Galaxies: elliptical and
  lenticular, cD -Galaxies: intergalactic medium }

\maketitle

\section{Introduction}

In recent years, intracluster stellar populations have been recognized as an important component of the
baryonic mass in clusters of galaxies. Their study can constrain their total baryonic budget, their interaction
history, and the enrichment of the intracluster medium (e.g. Gonzalez et
al. \cite{gonzalez07}; Gerhard et al. \cite{gerhard07}; Zaritsky et al. \cite{zaritsky04}).
Stellar populations visible as intracluster light (ICL) can contribute up to 50\% of the total stellar mass
(Lin and Mohr \cite{lin04}), although much lower numbers have also been
reported (Zibetti et al. \cite{zibetti05}; Krick \& Bernstein \cite{krick07}). In the present article, we focus on a phenomenon which is possibly related to the overall ICL, but much
more concentrated to the center of a galaxy cluster: the ``star pile'' in Abell 545.  

Abell 545 ($z=0.152$, Schneider et al. \cite{schneider83}) is one of the most
massive galaxy clusters known. With a richness class of 4, it is the
second richest in the Abell catalogue (Abell et al. \cite{abell89}). While
many clusters of this richness host a massive central cD galaxy,
Struble (\cite{struble88}) found on Palomar Sky Survey (POSS) plates in the center of Abell 545 a diffuse and very
 faint extended structure,
which he labelled ``star pile'', apparently a cD-halo without a cD galaxy.
Struble suggested that this was ``the first probable discovery of intracluster
matter'' in a rich cluster from visual inspection of the POSS, although he admitted the possibility that this
object may not belong to Abell 545, but instead being a Galactic nebula projected
onto Abell 545. However,  he also cites a private communication by J. Schombert, which suggested 
a reddish colour from a Palomar 5m uncalibrated CCD-image. Since Struble's paper there has
been no published follow-up work on this intriguing object. The paper has
been mentioned a few times, but until today, even the confirmation as an extragalactic
object is pending.

Assuming an association with Abell 545 (as we will confirm with this paper), the star pile may have significance in several
respects.  First, there is the question of its formation. Can tidal debris in
a very dense environment settle towards the cluster center? Is this
enlightening for the formation of cD halos? Then there  is the possibility
that such a structure may be used as a probe for the dark matter profile or
even for the very existence of dark matter. The CDM paradigm predicts very
dense and cuspy dark matter profiles. These cuspy profiles 
have however not been found in low luminosity spiral galaxies, nor in dwarf
galaxies (e.g. Gentile et
al.\cite{gentile04}; Simon et al. \cite{simon05}; Kuzio de Naray
et al. \cite{kuzio06}; Gilmore et al. \cite{gilmore07}).

 The centers of rich galaxy clusters should be the sites of the highest dark matter densities in the Universe, but
 are normally dominated by the baryons of a central cD-galaxy. A very faint
 baryonic structure such as the star pile might offer the possibility of
 distinguishing between dark matter and alternatives, such as MOND (Sanders \&
 McGaugh \cite{sanders02}). While this will be a challenging observational task for the future, the present
data are interesting enough to call the community's attention to this
extraordinary object.

In this contribution we present preliminary measurements of spectroscopic and
photometric properties of the star pile in Abell 545, which may provide new
insights to its origin. We
  assume throughout this paper a cosmology with $H_o=70$ km s$^{-1}$ Mpc$^{-1}$, $\Omega_M=0.3$ and $\Omega_{\Lambda}=0.7$.
This gives a distance modulus of $m-M = 39.3$.

\begin{figure}
\centering
{\includegraphics[width=0.46\textwidth]{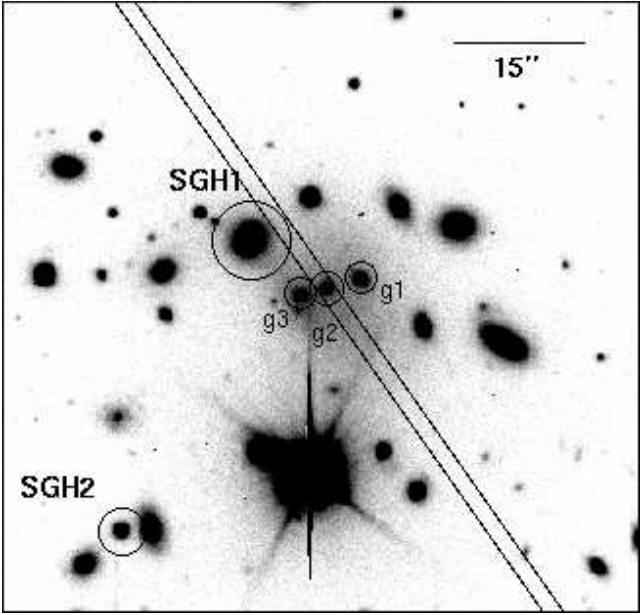}}
\caption{The image is an archival VLT image (see text) and shows the central region of Abell 545.
North is up, East to the left. The position of our long slit is indicated. The
star pile is visible as the diffuse elongated stucture with three faint
embedded sources (g1-g3). Note the bright star which complicates the
photometric analysis. Two galaxies (SGH1,2) are marked
for which Schneider et al. \cite{schneider83} already had given radial velocities. }
\label{image}
\end{figure} 

\section{Observations and analysis}

\subsection{Optical imaging}

The photometric data consist of 3 images in Bessell $V,R,I$ taken in the
photometric night 3/11/2002 and retrieved from the archive of the Very Large
Telescope (VLT) of the European Southern Observatory (programme 70.B-440(A),
PI L. Campusano), Cerro Paranal, Chile. The instrument was FORS1. The exposure time
for all three images was 460 sec. The image seeing values are 0.5\arcsec,0.59\arcsec,0.55\arcsec for $V,R,I$, respectively. 

Even though flatfielding using both sky-flats taken the same night
and ``master'' sky-flats provided by the observatory was applied, a still noticeable
gradient in the sky background remained. In order to minimize its effect
in the photometry, we determined the sky background in different blank sky areas
``close'' ($\sim$ 1 arcmin) to the star pile, but well outside its
apparent projected outer radius. Therefore  we assume  the uncertainty of
our sky determination to be  about 1\%, which  enters  the uncertainties of
the surface brightness profiles.

Better than on POSS-plates, the morphology of the star pile is discernable
on these VLT images (Fig.\ref{image}). One sees a diffuse elliptical structure
in which three faint sources are apparently embedded (already known to
Struble from  Schombert's CCD-image), the
middle one quite precisely in the geometrical center.

\begin{figure}
\centering
{\includegraphics[width=0.41\textwidth]{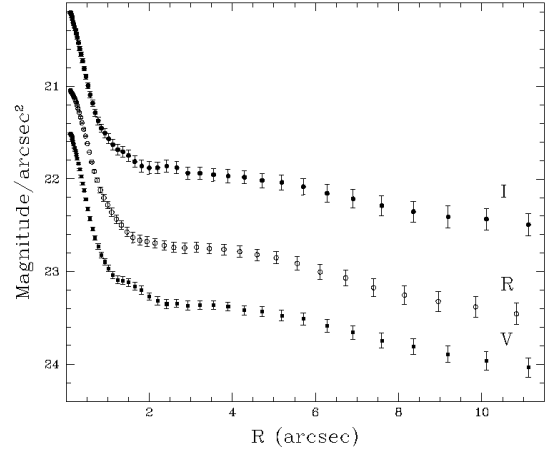}}
\caption{This graph shows the azimuthally averaged surface brightness profiles
  of the central galaxy plus the star pile in the bands Bessell
  $V,R,I$. At larger radii, the nearby bright star starts to dominate. At the distance of A545, one arcsec corresponds to 2.64 kpc. }
\label{profiles}
\end{figure}

The surface brightness profiles of the star pile in the bands $V,R,I$, derived from the archival images,
were obtained using the ELLIPSE package under IRAF, are shown in
Fig.\ref{profiles}.  Spikes from the nearby bright star and the
two neighbouring galaxies near the central one have been masked.
The absolute photometric calibration uses five stars from a standard field observed
the same night. The calibration equations obtained using PHOTCAL are:
\begin{eqnarray*}
\varv&=&V -27.43(\pm0.03) -0.039(\pm0.035)(V - I) +0.12X \\
r&=&R -27.34(\pm0.03) +0.021(\pm0.065)(V - R) +0.078X \\
i&=&I -26.52(\pm0.04) +0.043(\pm0.038)(V - I) +0.039X
\end{eqnarray*}
 with rms values of 0.03, 0.03 and 0.04, respectively, and where the extintion
 coefficients were obtained from the Paranal website. 

\begin{figure}
\centering
{\includegraphics[width=0.41\textwidth]{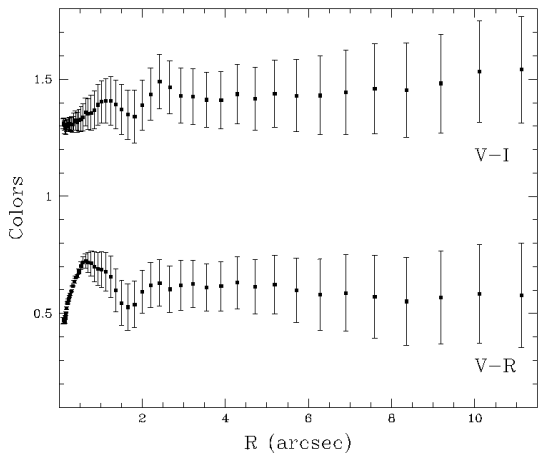}}
\caption{Colour profiles in Bessell $V-R$ and $V-I$ along the star pile's major axis. The innermost colour gradient is an effect of the different seeing of the images. One arcsec corresponds to 2.64 kpc.}
\label{colours}
\end{figure} 

The apparent magnitudes of the three embedded sources, measured through an
aperture of 1 arcsec, are  $V_{g1}=20.8, V_{g2}=21.9, V_{g3}=22.4$. These
numbers should be taken with caution, since the sky determination is
complicated by the star pile in which they are embedded. Nevertheless, the absolute magnitudes resemble
dwarf elliptical galaxies, although if they are considered as nuclei, their
absolute magnitudes should have been much brighter in the past. The colours are equal for all three
objects: $V-R = 0.7$ and $V-I = 1.3$. We adopt an absorption of 0.52 in $V$, and reddenings $E(V-R)=0.1$, and
$E(V-I) = 0.2$ (Schlegel et al. \cite{schlegel98}). That would let the colours be consistent with those of 
elliptical galaxies within the uncertainties.    

Fig.\ref{profiles} shows that the transition of the profile of the central
object occurs quite abruptly, which may be a morphological hint that the
central galaxy and the star pile are not associated. From Fig.\ref{colours}
one cannot derive convincingly a colour difference between the central object and the star pile and
consequently also not between the other objects and the star pile. 

\begin{figure}
\centering
{\includegraphics[height=0.35\textwidth,angle=-90]{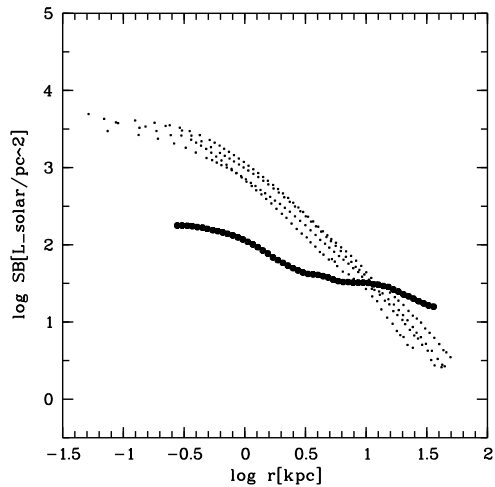}}
\caption{Comparison between the surface brightness profiles of the star pile and those of four
cD galaxies (Jord\'an et al. \cite{jordan04}) in units of $L_\odot/pc^2$ vs. kpc. Note that because of the logarithmic
scale, the major part of the star pile is in fact the central object and heavily influenced by the seeing.
The very shallow profile makes the star pile appear very different from the power-law cD galaxies.}
\label{cdgalaxies}
\end{figure}

Adopting a luminosity distance of 730 Mpc and an angular distance of 545 Mpc, we transform the surface brightness profile
into linear units of $L_{V,\odot}/pc^2$. To correct approximately for the restframe, we
further apply a factor $(1+z)^{-2}= 0.75$. The results are shown in Fig.\ref{cdgalaxies} together with the profiles
of four cD galaxies from Jord\'an et al. (\cite{jordan04}). It becomes very clear that the star pile is something different than a cD halo. The profile is so shallow that
also if the uncertainty of the sky level is taken into account, a typical cD power-law profile is excluded. 
 An excellent analytical description (without the claim for physical significance) for the profile is
$$
L_{V,\odot}/pc^2 = 154 \cdot (1+ (r/1.32)^2)^{-1.5} + 34.4 \cdot (1+(r/38.9)^2)^{-1.5}
$$
where $r$ is in kpc. 

Integrating this profile out to 30 kpc results in $M_V \sim -22.5$, resembling a
giant elliptical, but somewhat fainter than luminous cD-galaxies. The associated dynamical time, adopting
$M/L_V =8$, is, within 30 kpc, $1.6\times 10^8$ years. 

 Source  g1 in  Fig.\ref{image} is worth a particular remark. Colours of
$V-R = 0.7$ and $V-I = 1.3$ resemble an early type galaxy. We used ISHAPE (Larsen \cite{larsen99}) 
on the $V$-image (which has the best seeing) to find out whether
it is resolved. 
A King profile with a concentration parameter $c=100$, seems to be the best representation. It returns  an effective radius of only 0.8 pixels,
corresponding to about 400 pc.  
 With an absolute magnitude of $M_V = -18$ this source would 
resemble the very rare class of M32-like dwarf ellipticals, of which only a few representatives are known
(Mieske et al. \cite{mieske05}, Chilingarian et al. \cite{chilingarian07}). However, given that the resolution only is
marginal and that it may be influenced by the underlying star pile, we cannot
exclude with certainty that it is a  projected star, having by coincidence
similar colours as the star pile. Sources g2 and g3 have effective radii of
  1.55 kpc and 3.2 kpc, respectively.

\subsection{Long-slit spectroscopy}

The observations were carried out at Gemini South, Cerro Pachon, Chile, during the nights 19/22/29-11-2006, 25/27-12-2006,
and 20/21-1-2007
(programme GS-2006B-Q-69) in queue mode. The instrument was the Gemini Multi Object Spectrograph (GMOS-S).
Several long-slit spectra of the star pile were obtained along its major axis as indicated in Fig.\ref{image}.
  A total of
3$\times$ 2000 sec exposure time with central wavelength 5000\,\AA plus 3 $\times$ 1980 sec with central wavelength 4900~$\AA$ exposure was taken giving a
total of 3.3 hours exposure on-source. The grating B600\verb1+_1G5323
was used which provides a dispersion of 0.45 $\AA$/pixel. Since the star pile feature has such
low surface brightness we chose a slit width of 
2.0 arcsec to collect more light, which however degraded the FWHM of the arc lines to 
$\sim$ 9.5 $\AA$.  
To improve
S/N, but not degrade the spectral resolution, a spatial binning of $\times$ 4
was applied. The spectra cover the range 3700-6400 \AA.

The reduction of the spectra was done using the Gemini IRAF\footnote{IRAF is distributed by the
National Optical Astronomy Observatories, which are operated by the
Association of Universities for Research in Astronomy, Inc., under
cooperative agreement with the National Science Foundation.} package
(v1.9.1). Basic reduction (bias, flat fielding) was done in the standard way. 
The wavelength calibration was done with the standard Cu-Ar arc, using
GSWAVELENGTH, and left residuals with an rms-value of $\sim0.3 \AA$. The
2D spectra sharing the same central wavelength were
combined using GEMCOMBINE. These combined frames have been sky-subtracted
interactively, taking the sky at both sides of the star pile to avoid spatial
sky variations and at a distance of about 1\arcmin. Then the spectra were 
extracted and combined. 

\begin{figure}
\centering
{\includegraphics[width=0.48\textwidth]{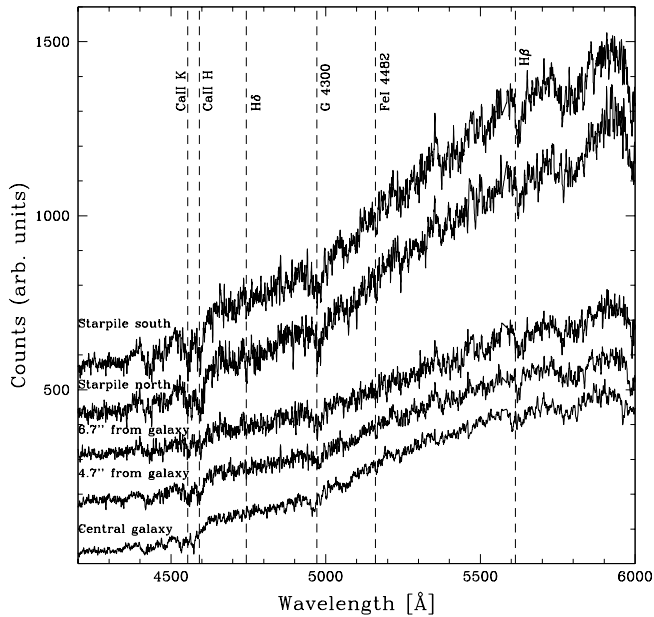}}
\caption{Spectra are shown of the central object and the star pile at two radii from the central galaxy as indicated in the figure. The uppermost two spectra of the star pile have been extracted over an extended radial range north and south of the central object. The spectra are smoothed by a median of 5 pixels. The star pile has a
velocity about 1300 km/s higher than the central galaxy. Dashed lines indicate
the positions of some spectral features in the star pile.} 
\label{spectra} 
\end{figure}

Fig.\ref{spectra} shows the spectra extracted at different  distances from the central object (g2). The lowest spectrum is that of g2 itself  and
has the usual appearance of an early-type galaxy.
 Radial velocities were obtained using both identification of lines (using the 
\verb-rvidlines- task) and cross-correlation (using \verb-fxcor-). In the latter case we used
K0III stars from the ELODIE stellar library~\footnote{http://www.obs.u-bordeaux1.fr/m2a/soubiran/elodie\_library.html}
 as
radial velocity templates.
  The heliocentric radial velocity of the star pile is 47100$\pm$60
kms$^{-1}$, the uncertainty resulting from the measurements of 16 well identifiable
lines. That corresponds to $z$= 0.157, thus confirming that the star pile is
indeed associated with Abell 545.

However, a striking feature
is that the star pile has a velocity of about 1300$\pm$ 80 km/s larger than the central object. No velocity differences among the
spectra of the star pile itself can be seen.
 The spectra of both the central galaxy and the star pile are too noisy to sensibly  measure line strengths, but the relatively weak Balmer lines indicate an old and metal-rich population.
The absence of a noticable 3727 $\AA$ emission line supports this in that there is no star formation apparent. Unfortunately, little can be said about the outermost spectrum.

\subsection{X-ray observations}

\begin{figure}
\centering
{\includegraphics[width=0.44\textwidth]{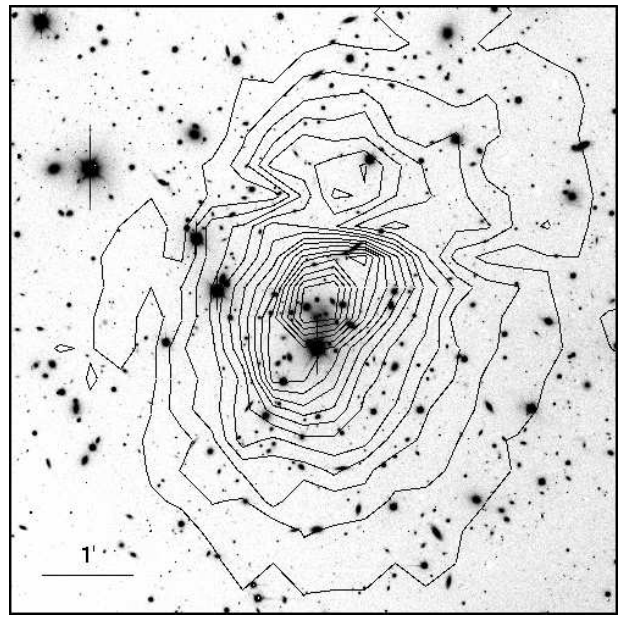}}
\caption{X-ray contour levels superimposed on the FORS1 $V$ image of the center
  of Abell 545. The star pile clearly lies at the center of the cluster.} 
\label{xrays} 
\end{figure}

Although Struble (\cite{struble88}) determined the position of the star pile
as being ``near'' the center of Abell 545 through galaxy counting, more precise information is of utter importance to unveil the origin of this feature.
We have retrieved an X-ray image of the cluster in the 0.5-2.0 KeV band from the XMM-Newton Science Archive
(PI: G. Madejski). Since at this stage we are only interested in the location
of the center of the cluster we simply generated a contour level plot, which
is seen superimposed to the FORS1 $V$-image in
Fig. \ref{xrays}. The image shows that the star pile is located in the center
of the cluster. The elongation of the contour levels have already been
detected on previous ROSAT images, and it may be an indication of dynamical
youth of the gas distribution (Buote \&
Tsai \cite{buote96}). Obviously a proper treatment of this data would reveal
more clues on the formation of the star pile, and it will be included in a
future paper.

\section{Discussion}

Schneider et al. (\cite{schneider83}) quote radial velocities for two of the galaxies marked in
Fig.\ref{image}. The velocity for SGH1 is 46200 km/s, for SGH2 48360 km/s. That means a difference between SGH2 and our
central galaxy of 2600 km/s, a value not uncommon for clusters as massive as
this one (Czoske \cite{czoske04}).
 The star pile's velocity (47100 km/s) is between these values,
consistent with its central position in the cluster.
 However, it is unknown, whether the star pile really is at rest at the dynamical center. Some other
cases have been found where the central cD galaxy has a significant peculiar velocity with respect
to the host cluster's rest frame, the largest one so far being 600 km/s (Pimbblet et al. \cite{pimbblet06}).
The radial velocity offset of 1300 km/s between the star pile and the central galaxy 
implies an even larger three dimensional velocity difference.  The central velocity dispersions
of massive clusters are of the order 1000-1500 km/s. Thus, unless the star pile itself has a comparable
velocity in the cluster's restframe, the central galaxy should have achieved its velocity by
travelling through a significant part of the cluster potential. In that case, the spatial association
with the star pile is plausible, but the dynamical association is doubtful. 

 Many cD galaxies are known to have
multiple nuclei (Gregorini et al. \cite{gregorini94}; Laine et al. \cite{laine03}). Perhaps we are witnessing an unusual case of galactic
cannibalism where strong tidal forces disrupted larger galaxies, leaving only the
dense nuclei intact (e.g. Gregg \& West \cite{gregg98}; Bekki et
al. \cite{bekki01}; Cypriano et
al. \cite{cypriano06}), or where totally disrupted dwarf galaxies have
become part of the intracluster medium (L\'opez-Cruz et
al. \cite{lopezcruz97}; Pracy et al. \cite{pracy04}).
There is also the possibility that multiple nuclei are
transient phenomena (Merritt \cite{merritt84}; Smith \cite{smith85}). 
A
scenario in which three or more elliptical galaxies interacted tidally, leaving the star pile as tidal debris and are
now visible as the two or three ``nuclei'' is perhaps imaginable, but must wait further data to get support. 

Although Rines et al. (\cite{rines07}) recently have discovered a similar
``star pile'' at a larger redshift in the cluster CL0958+4702, that kind of
event is unlikely to be the progenitor of the star pile in Abell 545, since the
galaxies interacting in CL0958+4702 would rapidily ($\sim 1$ Gyr) evolve into a
single brightest cluster galaxy (BCG) that is not seen in A545. The key difference
between both cases might be the mass of the cluster (A545 is more massive),
and hence the velocity dispersion. In extremely massive clusters the
formation of a BCG might be prevented due to the total disruption of the
interacting galaxies.

As Fig.\ref{cdgalaxies} shows, the difference between the shallow profile of the star pile and that of cD-galaxies
is striking. Often, cD galaxies are characterized as ``normal'' galaxies with an additional cD-halo
which is manifest by the deviation from a de Vaucouleur profile. But apparently, their profiles can
be better described by power laws with an exponent near -3, in which case the cD-halo is not
identifiable as a distinct morphological entity. This is so for NGC 1399 (Dirsch et al. \cite{dirsch03}),
 NGC 6166 (Kelson et al. \cite{kelson02}) and
also for the four galaxies in Fig.\ref{cdgalaxies}.
It is interesting that we can also fit the profile of the star pile by such a power law,
but with a core of the size of its visible extension. 
 So one may ask, whether the star pile is only the brightest part of a much more extended
structure. In that case it should be dynamically at rest, at the center of the
cluster. A significant non-zero velocity in the
cluster's rest frame would disprove this possibility immediately.

Finally, we point out an interesting possibility of using the star pile in
Abell 545 to trace the dark matter distribution in the center of a massive
galaxy cluster as a test of the standard CDM paradigm. One of the predictions
of the CDM paradigm is that dark matter halos should develop cuspy central
density profiles.
Sites with a high ratio of dark matter density to baryonic density are for example low surface brightness
galaxies, where, however, these cuspy profiles have not been found (e.g. Gentile et al. \cite{gentile04}). Instead, the dynamics seem to
be consistent with Modified Newtonian Dynamics (Gentile et al. \cite{gentile07}). The centers of massive galaxy clusters should exhibit the
highest dark matter densities  in the Universe, but if there is a stellar tracer population then it is usually
a massive cD galaxy which then baryonically  dominates the total density. The
star pile, if it is at the dynamical center, has a low central density, much
more suitable for tracing the dark matter. This, of course, is only viable if the star pile is in equilibrium. 
X-ray studies are in principle equally suited, but departures from hydrostatic equilibrium are probably
most pronounced near the very center of a galaxy cluster (Ciotti \& Pellegrini
\cite{ciotti04}; Diehl \& Statler \cite{diehl07}).

\begin{figure}
\centering
{\includegraphics[width=0.35\textwidth,angle=-90]{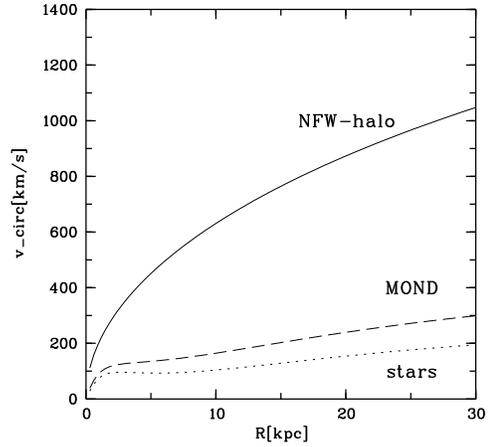}}
\caption{This graph shows three hypothetical circular velocity curves for the star pile. The lowest curve refers to the stellar population only, the middle one assumes MOND and the uppermost an NFW-halo with parameters as given in the text.}
\label{vcirc}
\end{figure} 

We illustrate a possible
situation in Fig.\ref{vcirc}, where we draw the expected circular velocities for the different components.
The uppermost curve describes an NFW-halo with a characteristic radius of 300
kpc and a mass of $1.6\times 10^{15} M_\odot$.
We motivate this mass for Abell 545 by applying the relation between total X-ray luminosity and galaxy cluster mass
of Stanek et al. (\cite{stanek06})  and an
 X-ray luminosity of $3.5\times 10^{44}$ erg/s, quoted by B\"ohringer et al. (\cite{boehringer04}).

 The lowest curve represents the star pile after deprojecting the profile of Fig.\ref{profiles} and assigning an
$M/L_V$-value of 7. The observable projected velocity dispersions will be somewhat lower, but since the shallow
 profile enhances the projected dispersion with respect to a ``normal''
 power-law galaxy, it should be less than a factor of about 1.5. Although it will
 be difficult to measure a velocity dispersion profile for such a faint
 structure, it is not impossible.

\section{Conclusions}
The star pile belongs to Abell 545 and is located in the center of the cluster. The spectrum indicates an old, probably metal-rich
population, which would suggest an origin from tidal debris rather than from primordial intergalactic material. 
 The central object is dynamically decoupled and probably does not play the role of a nucleus
in a cD-halo. The surface brightness profile is much too shallow to represent a cD galaxy, although
the total brightness of the star pile is of the order of what one would expect for a giant elliptical. 
A possible scenario is that the three embedded sources interacted tidally, leaving the star
pile as tidal debris. More insight into this remarkable object must come from high S/N-spectra of the
star pile in combination with a dynamical analyis of the cluster.

\begin{acknowledgements}
We thank an anonymous referee for her/his helpful comments, which 
improved the paper.
We thank Sonia Duffau, Alastair Edge and Mat\'ias G\'omez for helpful discussions.
R.S. gratefully
acknowledges support from a CONICYT Doctoral Fellowship. T.R. and A.J.R. gratefully
acknowledge support from the Chilean Center for Astrophysics, FONDAP
No. 15010003. A.J.R. also acknowledges support by NSF grant 
AST-0507729.

\end{acknowledgements}

\bibliographystyle{aa}
\end{document}